\documentclass{pasj00}

\draft

\begin{document}
\SetRunningHead{S. Kato}{}
\Received{2009/00/00}%{yyyy/mm/dd}
\Accepted{2009/00/00}%{yyyy/mm/dd}

\title{Resonant Excitation of Disk Oscillations 
in Two-armed-Deformed Disks and Application to High-Frequency QPOs}

%%% begin:list of authors
\author{Shoji \textsc{Kato}}
    %\thanks{Example: Present Address is xxxxxxxxxx}}
\affil{}
\email{kato@gmail.com, kato@kusastro.kyoto-u.ac.jp}
%\and
%\author{C-Firstname {\sc C-Familyname}}
%\affil{C-Address of Institute}\email{ccccc@xxx.xxx.xx.xx}
%%% end:list of authors

%%% Please use the following style in case that sorting by 
%%% affilation is impossible. 
%
% \author{%
%   D-Firstname \textsc{D-Familyname}\altaffilmark{1}
%   E-Firstname \textsc{E-Familyname}\altaffilmark{1,2}
% and
%   F-Firstname \textsc{F-Familyname}\altaffilmark{2}}
% \altaffiltext{1}{Address of Institute}
% \email{ddddd@xxx.xxx.xx.xx}
% \email{eeeee@xxx.xxx.xx.xx}
% \altaffiltext{2}{Address of Institute}

%% `\KeyWords{}' always has to be placed before `\maketitle'.
 %Do NOT move this preamble from here!

\maketitle

\begin{abstract}
In previous papers we showed that in a one-armed deformed disks, p-mode and g-mode
oscillations are resonantly excited by horizontal resonance, and applied it to 
high frequency QPOs observed in low mass X-ray binaries.
In that model, the observed time variation of kHz QPOs is regarded as a result of 
a time-dependent precession of the deformation.
In this paper we consider another possible cause of time variation of kHz QPOs.
That is, we demonstrate that in a two-armed deformed disks, p-mode and g-mode
oscillations are excited by {\it vertical resonance}, not by horizontal resonance
(horizontal resonance dampens them).
Furthermore, we show that in the case of vertical resonance, the frequencies of disk 
oscillations excited can vary
with time if vertical disk structure changes with time.
A brief application of these results to the time variation of observed kHz QPOs is made. 

\end{abstract}

\section{Introduction}

Many authors now think that high frequency quasi-periodic oscillations (QPOs)
observed in low mass X-ray binaries are some kind of disk oscillations in strong 
gravitational field, and that it gives a promising way to estimate the mass and spin of the
central sources.
High frequency QPOs both in black hole candidates and in neutron star sources will have the same dynamical 
origin (Abramowicz et al. 2003), although there are some differences in their frequency variations.

Based on closeness to 3 : 2 of frequencies of twin QPOs, importance of resonant phenomena was
emphasized by Abramowicz and Klu\'{z}niak (2001).
Subsequently, from a different context, Kato (2004, 2008a,c) pointed out the importance of resonant processes
in a deformed (warped or eccentric) disks as an excitation process of high frequency QPOs.
That is, 
%in In previous papers we examined resonant excitation of disk oscillations 
%in one-armed (warped or eccentric) disks in order to
%suggest a possible origin of high frequency QPOs in low mass X-ray binaries.
%We showed (Kato 2004, 2008a,c) that 
non-linear interaction between disk oscillations and
the deformed part of disks resonantly excite or dampen disk oscillations.
There are two types of the resonance.
One is a horizontal resonance (Lindblad resonance) and the other is a vertical resonance.
The inertial-acoustic oscillations (p-mode) and the gravity oscillations (g-mode)
are found to be excited by the horizontal resonance.

The resonant excitation of the p-mode and g-mode oscillations by the horizontal resonance 
seems to occur most efficiently at the radius where the condition
\begin{equation}
      \kappa=\frac{\Omega}{2}
\label{1.1}
\end{equation}
is satisfied (Kato 2004, 2008a,c), when the one-armed deformation of disks have no precession,
where $\kappa(r)$ is the epicyclic frequency, $\Omega(r)$ is the angular velocity
of disk rotation and $r$ is the radius from the central source on the disk plane.
In the case of Schwarzschild metric, the resonance occurs at $4r_{\rm g}$, where
$r_{\rm g}$ is the Schwarzschild radius defined by $r_{\rm g}=2GM/c^2$,
$M$ being the mass of the central source.
The frequencies of the p-mode and g-mode oscillations excited by this resonance are
\begin{equation}
    \omega=(m\Omega\pm\kappa)_{\rm res},
\label{1.2}
\end{equation}
where the subscript res denotes the values at the resonant radius derived from equation (\ref{1.1})
and $m(=1,2,...)$ are the wavenumber of the oscillations in the azimuthal direction.
The frequencies specified by equation (\ref{1.2}) have ratios of some rational numbers.

The frequencies of oscillations given by equation (\ref{1.2}) can qualitatively describe the
high frequency QPOs in black-hole and neutron-star X-ray binaries.
In the case of neutron-star X-ray binaries, however, there is an additional and challenging 
observational evidence that the frequencies of kHz QPOs change with time. 
If we want to describe this time change of kHz QPOs in the framework of the above-mentioned model,
we must introduce the assumption that the one-armed disk deformation has time-dependent
precession in the disks of neutron-star X-ray binaries\footnote{
In the case of neutron stars, the central sources have surfaces and the inner part of disks may have 
influences of stellar magnetosphere. 
In such situations, the warp might have precession (e.g., Meheut and Tagger 2009). 
}
, although such precession is not required in the
black-hole accretion disks.
If the frequency of precession of disk deformation is denoted by $\omega_{\rm p}$, the resonant condition
where the horizontal resonance occurs is changed from equation (\ref{1.1}) to
\begin{equation}
      \kappa=\frac{1}{2}(\Omega-\omega_{\rm p}).
\label{1.3}
\end{equation}
The resonant radius, $r_{\rm res}$, changes with a time change of $\omega_{\rm p}$, and thus
the frequencies of resonantly excited disk oscillations given by equation (\ref{1.2}) changes with time
with correlation.

This may be one of causes of the time variation of kHz QPOs.
However, a question naturally raised here is whether a variation of precession frequency that
is enough to account for the time variation of kHz QPOs is
really expected in disks of neutron-star X-ray binaries. 
Hence, it will be worthwhile considering other possibilities of frequency change of kHz QPOs.
This is the purpose of this paper.

First, we should notice that in the case of vertical resonance, possibility of frequency variation
of resonant oscillations is high compared with in the case of horizontal resonance. 
This is because the radius where the vertical resonance occurs is sensitive to the vertical structure
of disks\footnote{
In the horizontal resonance, the resonant radius is determined by the radial distributions of the 
Keplerian angular velocity and the radial epicyclic frequency, and they are almost time-independent in
geometrically thin disks.
}.
In spite of this interesting characteristics of the vertical resonance (Kato 2005), we already showed that the disk 
oscillations resulting from vertical 
resonance are damped, not excited, in the case where the disk deformation is one-armed (Kato 2004, 2008a,c).

In the case where the disk deformation is two-armed, however, the situations are changed.
That is, as is suggested in this paper, the resonance that can excite 
the p-mode and g-mode oscillations is not horizontal one, but vertical one.
This means that if the disk deformation in neutron-star X-ray binaries is two-armed, different from the case of
black-hole X-ray binaries,
the observed frequency variations of kHz QPOs will be naturally accounted for by taking the viewpoint 
that kHz QPOs are p- and g-mode oscillations resonantly excited by vertical resonance.
 
In section 2, we briefly summarize the essence of resonant excitation of disk oscillations in one-armed deformed
disks as a preparation for examining the case where the disk deformation is two-armed.
Resonant excitation of p-mode and g-mode oscillations by vertical resonance in vertically isothermal disks
is examined in section 3.
In section 4, the results in section 3 are extended to the case of vertically polytropic disks, and 
time variation of excited oscillations by a change of vertical disk structure is discussed.
Numerical results are given in section 5, and the final section is devoted to discussion. 

\section{Brief Summary of Resonant Excitation of Disk Oscillations in Warped Disks}

Before considering resonant excitation of disk oscillations in two-armed deformed-disks, 
we shall briefly summarize the essence of resonant excitation of disk oscillations in warped disks,
since the procedure for examining the former problem is a simple modification of the latter.

\subsection{Unperturbed Disks}

For mathematical simplicity, the unperturbed disks are assumed to be isothermal in the vertical
direction.
This leads to the fact that the density is stratified exponentially in the vertical direction as
\begin{equation}
    \rho_0(r,z)=\rho_{00}(r){\rm exp}\biggr[-\frac{z^2}{2H^2(r)}\biggr],
\label{2.1}
\end{equation}
where $\rho_{00}$ is the density on the equatorial plane ($z=0$), and
$H(r)$ is the half-thickness of the disk, $r$ being the distance from the rotating axis of the disks.

The half-thickness, $H$, of disks is related to the vertical epicyclic frequency, $\Omega_\bot(r)$, by
\begin{equation}
      \Omega_\bot^2H^2=\frac{p_{00}}{\rho_{00}}=c_{\rm s}^2(r),
\label{2.2}
\end{equation}
where $c_{\rm s}$ is the isothermal acoustic speed.
The vertical epicyclic frequency, $\Omega_{\bot}(r)$, is equal to the angular velocity of the Keplerian 
rotation, $\Omega_{\rm K}(r)$, in the case of non-rotating central object, and practically
equal to the angular velocity of disk rotation, $\Omega(r)$.
Hereafter, however, we do not use $\Omega_{\rm K}$ or $\Omega$ instead of $\Omega_\bot$ 
so that we can trace back the effects of $\Omega_\bot$ on the final results.

\subsection{One-armed Disk Deformation}

We assume that the disks described above are deformed from axisymmetric state by some external
or internal cause.
The deformation is a warp or an eccentric deformation in the equatorial plane.
They are assumed, for simplicity, to be time-independent.

The Lagrangian displacement associated with the deformation, $\mbox{\boldmath $\xi$}^{\rm W}
(\mbox{\boldmath $r$})$, is denoted by
\begin{eqnarray}
  &&\xi^{\rm W}_r={\rm exp}(-i\varphi)\hat{\xi}_r^{\rm W}(r){\cal H}_{n^{\rm W}}(z/H),
                                  \nonumber \\
  &&\xi^{\rm W}_\varphi={\rm exp}(-i\varphi)\hat{\xi}_\varphi^{\rm W}(r){\cal H}_{n^{\rm W}}(z/H),
                                  \nonumber \\
  &&\xi^{\rm W}_z={\rm exp}(-i\varphi)\hat{\xi}_z^{\rm W}(r){\cal H}_{n^{\rm W}-1}(z/H),
\label{2.3}
\end{eqnarray}
where $\varphi$ is the azimuthal direction of the cylindrical coordinates ($r$, $\varphi$, $z$),
${\cal H}_{n^{\rm W}}$ is the Hermite polynomial of order $n^{\rm W}$ with
argument $z/H$.
In the case of eccentric deformation in the equatorial plane, $n^{\rm W}$ is zero
($n^{\rm W}=0$), while it is unity ($n^{\rm W}=1$) in the case of warp.
It is noted that the number of nodes of $\xi_z^{\rm W}$ is smaller than those of $\xi_r^{\rm W}$ 
and $\xi_\varphi^{\rm W}$ by one.
This is true even for disk oscillations [see equation (\ref{2.4})].

\subsection{Disk Oscillations}

Disk oscillations that are considered here are assumed to have moderately short radial wavelength
so that it is shorter than the
characteristic length of radial variation of disk structure.
Furthermore, the wave motions are assumed to occur isothermally.

If the above approximations are adopted, the $r$- and $z$- dependences of disk oscillations 
on non-deformed disks are approximately separated (Okazaki et al. 1987).
We can then express the displacement vector, $\mbox{\boldmath $\xi$}(\mbox{\boldmath $r$},t)$,
associated with an oscillation mode of ($\omega$, $m$, $n$) as
\begin{eqnarray}
    &&\xi_r(\mbox{\boldmath $r$},t)={\rm exp}[i(\omega t-m\varphi)]
              \hat {\xi}_{r,n}(r){\cal H}_n(z/H),   \nonumber \\
    &&\xi_\varphi(\mbox{\boldmath $r$},t)={\rm exp}[i(\omega t-m\varphi)]
              \hat{\xi}_{\varphi,n}(r){\cal H}_n(z/H),  \nonumber \\
    &&\xi_z(\mbox{\boldmath $r$},t)={\rm exp}[i(\omega t-m\varphi)]
              \hat {\xi}_{z,n}(r){\cal H}_{n-1}(z/H),
\label{2.4}
\end{eqnarray}
where $\omega$ is the frequency of the oscillations,  and ${\cal H}_n$ is the Hermite polynomials
as mentioned before.
The integer $n (=0,1,2,...$) specifies the number of nodes of $\xi_r$ (and $\xi_\varphi$) in the
vertical direction ($z$-direction).

Hereafter, for simplicity, we characterize the oscillations and disk deformations by the set of 
($\omega$, $m$, $n$).
For example, a warp belongs to a mode characterized by (0,1,1) and an eccentric deformation on 
the disk plane by (0,1,0).

Here, classification of disk oscillation modes is briefly summarized (for details, see Kato 2001;
Kato et al. 2008).
The oscillations with $n=0$ occur predominantly on the equatorial plane.
They are inertial-acoustic oscillations and called hereafter "p-mode" oscillations.
In the cases where $n\geq 1$, we have two kinds of oscillations.
In one of them, $(\omega-m\Omega)^2<\kappa^2$, while in the other one we have $(\omega-m\Omega)^2>
n\Omega_\bot^2$.
We call the former "g-modes", and the latter "vertical p-modes" except for some special cases mentioned below.
In the case of $n=1$ with $m=1$ (and some special cases of $n\geq 2$ with $m\geq 2$), 
the latter oscillations are almost incompressible and have low frequencies.
They are specially called "c-mode" oscillations.
A warp is included to this type of oscillations.

In the resonant excitation problem to be examined in this paper, we group the above various oscillation modes into
two classes and treat the oscillation modes in each class as a pack, due to similarity of mathematical treatment.
One class is p-mode and g-mode oscillations, and the other one is vertical p-mode and c-mode oscillations.

\subsection{Coupling between Deformation and Oscillations}

Nonlinear couplings between the deformation specified by equations (\ref{2.3}) and the disk oscillations 
given by equations (\ref{2.4}) induce disk oscillations that are characterized by 
($\omega$, $m\pm 1$, ${\tilde n}$),
where ${\tilde n}=n\pm 1$ when the disk deformation is a warp, while ${\tilde n}=n$ when the deformation 
is an eccentric deformation on the equatorial plane.
Arbitrary combinations of $\pm$ are allowed.
We call the oscillations resulting from the coupling "intermediate oscillations".

The intermediate oscillations have resonant interaction with the disks at particular radii.
One of resonances occurs at the radii where the intermediate oscillations with ($\omega$, $m\pm 1$, ${\tilde n}$) 
have the Lindblad resonances, which are specified by 
\begin{equation}
    [\omega-(m\pm 1)\Omega]^2-\kappa^2=0,
\label{2.5}
\end{equation}
where $\kappa$ is the (horizontal) epicyclic frequency.
This resonance is call hereafter "horizontal resonance".
Another one occurs at the radii where the frequency $\omega$ of the intermediate oscillations of 
($\omega$, $m\pm 1$, ${\tilde n}$) becomes equal to the eigen-frequency of vertical
oscillations of disks.
The radii are characterized by
\begin{equation}
   [\omega-(m\pm 1)\Omega]^2-{\tilde n}\Omega_\bot^2=0.
\label{2.6}
\end{equation}
This resonance is callded hereafter "vertical resonance".

\subsection{Resonant Excitation of Oscillations}

The intermediate oscillations interact nonlinearly with the disk deformation, after 
having the resonance mentioned above, to feedback to the original oscillations.  
This feedback process amplifies or dampens the original oscillations.
Detailed analyses (Kato 2004, 2008a,c) show that in both cases of horizontal and
vertical resonances, the growth rate of oscillations, $-\omega_{\rm i}$ ($\omega_{\rm i}$ 
being the imaginary part of frequency of oscillations), can be expressed in the form of
\begin{equation}
  -\omega_{\rm i}\propto \frac{{\rm sign}[\omega-(m\pm 1)\Omega]_{\rm res}}{E},
\label{2.7}
\end{equation}
where $E$ is the wave energy of the original oscillations ($\omega$, $m$, $n$) in consideration, and
${\rm sign}[\omega-(m\pm 1)\Omega]_{\rm res}$ is the sign of $\omega-(m\pm 1)\Omega$ at the resonant
radius, $r_{\rm res}$.\footnote{
The resonant radius, $r_{\rm res}$, is not uniquely determined by the resonant condition (\ref{2.5}) or (\ref{2.6})
alone.
An additional condition is necessary, which will be discussed later in subsection 2.6.
}
The value of the proportional coefficient of equation (\ref{2.7}) depends, of course, on modes
of oscillations and types of resonances.
However, in some typical cases, the coefficient is always negative definite (Kato 2004, 2008a,c).
This means that the condition of excitation of disk oscillations is
\begin{equation}
     \frac{{\rm sign}[\omega-(m\pm 1)\Omega]_{\rm res}}{E}<0.
\label{2.8}
\end{equation}

This condition allows us to have a simple physical interpretation.
We notice first that for a resonance to occur efficiently, the resonant radius, $r_{\rm res}$, must be in the 
radial region where both the original and the intermediate oscillations are dominated.
Furthermore, we notice that in general a wave with ($\omega$, $m$) has a negative energy if
the wave is dominated inside the corotation radius given by $\omega-m\Omega=0$, while it is positive
if the wave is outside the corotation radius.
This consideration suggests that we can regard the sign$[\omega-(m\pm 1)\Omega]_{\rm res}$ as the sign
of the wave energy, $E^{\rm int}$, of the intermediate oscillation.
Furthermore, the sign of the wave energy of the original oscillation is the same as 
the sign of $(\omega-m\Omega)_{\rm res}$.
Based on these considerations, we can write the amplification condition (\ref{2.8}) as
\begin{equation}
       \frac{E^{\rm int}}{E} <0 
\label{2.9a}
\end{equation}
or 
\begin{equation}
      \frac{{\rm sign}[\omega-(m\pm 1)\Omega]_{\rm res}}{{\rm sign}(\omega-m\Omega)_{\rm res}}<0.
\label{2.9b}
\end{equation}

This condition can be interpreted in the following way.
If an oscillation with positive energy ($E>0$) resonantly interacts, at a resonant radius, 
with an intermediate oscillation
with negative energy ($E^{\rm int}<0$), both oscillations are
amplified by energy flowing from the intermediate oscillation to the original oscillation.
The original oscillation is amplified, since it has $E>0$ and receives positive energy.
The intermediate oscillation also grows by loosing energy since $E^{\rm int}<0$.
In the case of $E<0$ and $E^{\rm int}>0$, the resonance also amplifies the oscillations.
In this case, the direction of energy flow is opposite:
It flows from the original oscillation to the intermediate oscillation at the resonant
radius.

\subsection{Resonant Radius and Type of Resonance That Excites Oscillations}

It is noted that the resonant condition [equation (\ref{2.5}) or (\ref{2.6})] alone
does not uniquely determine the radius of resonance.
An additional restriction is necessary.
We assume that the original oscillations are most strongly excited when the resonance
occurs near the boundary of their propagation region.
Near the boundary the oscillations have long radial wavelength and their group velocity
in the radial direction vanishes, 
i.e., they stay there for a long time compared with in other places, and thus will grow there
most strongly.
Hence, when we consider p- and g- mode oscillations, we assume that an additional condition
to be adopted to specify the resonant radius is 
\begin{equation}
     (\omega-m\Omega)^2-\kappa^2=0,
\label{2.10}
\end{equation}
since this represents the boundary of the propagation region of p- and g-mode oscillations.
In the case where we consider vertical p-mode oscillations and c-mode oscillations, 
on the other hand, we assume that the resonance occurs at
\begin{equation}
       (\omega-m\Omega)^2-n\Omega_\bot^2=0,
\label{2.11}
\end{equation}
by the same reason as the above.

As mentioned in subsection 2.3, disk oscillations can be grouped into two classes in studying the present
excitation problem.
The first one is the p- and g-mode oscillations.
The second class is the vertical p-mode and c-mode oscillations.
Concerning the type of resonances, we have also two types, i.e., the horizontal and
vertical resonances.
Hence, we have four cases in combination of the set of oscillations and resonances.
Among them, the excitation of disk oscillations occurs in the case where 
the oscillations are p- and g-modes and the resonance is horizontal (Kato 2004, 2008a, c). 
Combining equations (\ref{2.5}) and (\ref{2.10}), we find that the resonance in this
growing case occurs at the radius where the condition of
\begin{equation}
       \kappa=\frac{1}{2}\Omega
\label{2.12}
\end{equation}
is satisfied.
The radius where this condition is satisfied is $4r_{\rm g}$, i.e., $r_{\rm res}=4r_{\rm g}$,
when the metric is the Schwarzschild one.
If the metric is the Kerr, the radius becomes smaller than $4r_{\rm g}$.

The frequencies of the disk oscillations that are excited there are [see equation (\ref{2.10})]
\begin{equation}
      \omega=(m\Omega\pm \kappa)_{\rm res}.
\label{2.13}
\end{equation}
Application of these resonantly-excited oscillations to high-frequency QPOs is made by
Kato and Fukue (2006) and Kato (2008b).

\section{Resonant Excitation of Disk Oscillations in Two-Armed Disks}

After the above preparation, we now proceed to the main purpose of this paper,
i.e., examination of the case where the disks are deformed from the axisymmetric steady state into
a state with a two-armed pattern.
Different from the case of a one-armed pattern, a two-armed pattern cannot be stationary 
in general, i.e., it will be time-dependent, and wavy.
What we need here is that a pattern has approximately a constant frequency for a time
interval longer than the characterisitic time by which disk oscillations grow.
The origin of such a two-armed pattern is a problem to be discussed and clarified, but
here we simply assume that such a pattern exists on the disks by some internal or external causes.
For example, numerical 3D MHD simulations of accretion disks
(Machida and Matsumoto 2008) show that
one-armed and two-armed patterns with slow rotation are produced in the innermost 
region of disks at a certain stage of disk evolution by
magnetic-field streching and reconnection processes.

The displacement vector, $\mbox{\boldmath $\xi$}^{\rm T}(\mbox{\boldmath $r$},t)$, associated with
a two-armed deformation is now expressed as
\begin{eqnarray}
     &&\xi_r^{\rm T}(\mbox{\boldmath $r$},t)={\rm exp}[i(\omega^{\rm T}t-2\varphi)]\hat{\xi}_r^{\rm T}(r)
              {\cal H}_{n^{\rm T}}(z/H),     \nonumber \\
     &&\xi_\varphi^{\rm T}(\mbox{\boldmath $r$},t)={\rm exp}[i(\omega^{\rm T}t-2\varphi)]\hat{\xi}_\varphi^{\rm T}(r)
              {\cal H}_{n^{\rm T}}(z/H),     \nonumber \\         
     &&\xi_z^{\rm T}(\mbox{\boldmath $r$},t)={\rm exp}[i(\omega^{\rm T}t-2\varphi)]\hat{\xi}_z^{\rm T}(r)
              {\cal H}_{n^{\rm T}-1}(z/H).
\label{3.1}
\end{eqnarray}
Here, $\omega^{\rm T}$ is the angular velocity of the pattern and taken to be a free parameter.
As the integer $n^{\rm T}$, we are mainly interested in the cases of $n^{\rm T}=2$ and 3 (see section 6).

On such deformed disks, we impose disk oscillations.
The displacement vector, $\mbox{\boldmath $\xi$}(\mbox{\boldmath $r$},t)$, associated with the oscillations 
is described again by equation (\ref{2.4}).

\subsection{Resonant Conditions}

The nonlinear coupling between the disk deformation described by ($\omega^{\rm T}$, 2, $n^{\rm T}$) 
[see equation (\ref{3.1})] and
the disk oscillations described by ($\omega$, $m$, $n$) [see equation (\ref{2.4})] induces
the intermediate oscillations of ($\omega\pm\omega^{\rm T}$, $m\pm 2$, $\tilde {n}$), where
${\tilde n}=n\pm n^{\rm T}$.
These intermediate oscillations have resonances with the disk rotation.
The horizontal resonance occurs at [cf., equation (\ref{2.5})]
\begin{equation}
    [\omega\pm\omega^{\rm T}-(m\pm 2)\Omega]^2-\kappa^2=0,
\label{3.2}
\end{equation}
and the vertical resonance occurs at [cf., equation (\ref{2.6})]
\begin{equation}
    [\omega\pm\omega^{\rm T}-(m\pm 2)\Omega]^2-{\tilde n}\Omega_\bot^2=0.
\label{3.3}
\end{equation}

As mentioned in subsection 2.6, an additional condition is necessary to determine uniquely 
the radius of resonance.
We focus our attention again to the case where the resonance occurs at the
radius where the group velocity of the original oscillations with ($\omega$, $m$, $n$) vanished 
to stay there for a long time.
That is, when we consider p- and g-mode oscillations, we adopt [equation (\ref{2.10})] 
\begin{equation}
     (\omega-m\Omega)^2-\kappa^2=0,
\label{3.4}
\end{equation}
as the additional condition.
In the case where the excitation of vertical p-mode and c-mode oscillations is examined, we adopt 
[equation (\ref{2.11})]
\begin{equation}
      (\omega-m\Omega)^2-n\Omega_\bot^2=0.
\label{3.5}
\end{equation}

There are four cases in combination of types of resonance and types of oscillation. 
i.e., two cases (horizontal or vertical resonance) for p- and g-mode
oscillations, and two cases (horizontal or vertical resonance) for vertical 
p-mode and c-mode oscillations.
Before examining these cases separately in subsection 3.3, we consider the excitation condition.

\subsection{Excitation Condition}

In the case of one-armed deformation of disks, the condition of 
excitation of disk oscillations is given by equation (\ref{2.7}).
The mathematical procedures to derive the condition are complicated (see Kato 2008a, c), but
they can be straightly extended to the case of two-armed deformation.
The results show that the growth rate of oscillations, $-\omega_{\rm i}$, can be expressed in
the form of
\begin{equation}
    -\omega_{\rm i}\propto \frac{{\rm sign}[\omega\pm\omega_{\rm p}-(m\pm 2)\Omega]_{\rm res}}
         {E}.
\label{3.6}
\end{equation}
That is, compared with the case of the one-armed deformation, in the present
case of two-armed deformation, $\omega$ is changed to $\omega\pm\omega_{\rm p}$ and 
$m\pm 1$ is changed to $m\pm 2$.

The next problem is to examine the sign of the proportional coefficient on 
the right hand side of equation (\ref{3.6}).
A straightforward generalization of the procedures of the one-armed deformation suggests that
in some simplified cases 
(for example, the case where the non-linear coupling terms between the original oscillation and the
deformation are
constant in the resonant region\footnote{
The resonant region has a finite width in the radial direction around the resonant radius.
The width depends on the disk temperature.
In a pressureless disks, the width is infinitesimally narrow, but increases with increase of
disk temperature (e.g., Kato 2008c).
}), 
the proportional coefficient is negative definite as in the
case of one-armed deformation.
When the proportional coefficient is negative definite, we have again a simple physical 
interpretation of equation (\ref{3.6}) [see subsection 2.5].
Considering them we suppose that the amplification condition in the present two-armed case
is given by
\begin{equation}
        \frac{{\rm sign}[\omega\pm\omega_{\rm p}-(m\pm 2)\Omega]_{\rm res}}
         {{\rm sign}(\omega-m\Omega)_{\rm res}}<0.
\label{3.7}
\end{equation}
In other words, the condition is also expressed as $E^{\rm int}/E<0$, and has a simple
physical meaning.

\subsection{Growing Cases and Their Resonant Radius} 

Based on the resonant conditions [relevant combinations of equation {(\ref{3.2}) or (\ref{3.3})
to equation (\ref{3.4}) or (\ref{3.5})] 
and the excitation  condition [equation (\ref{3.7})], 
we now examine what types of oscillations (p- and g-mode oscillations or
vertical p- and c-mode oscillations) are excited by what type of resonances (horizontal or
vertical resonance) and where the radii are.

\bigskip\noindent
i) Horizontal resonance of p- or g-mode oscillations

In this case, as mentioned in subsection 3.1, the resonant radii are the places where both equations 
(\ref{3.2}) and (\ref{3.4}) are simultaneously satisfied.
More explicitely, horizontal resonances of p- and g-mode oscillations occur at the radii where one of the
following set of two equations are satisfied:
\begin{eqnarray}
     &&{\rm (a)}:\quad \omega\pm\omega^{\rm T}-(m\pm2)\Omega=\kappa \quad {\rm and}\quad
               \omega-m\Omega=\kappa,   \nonumber \\
     &&{\rm (b)}:\quad \omega\pm\omega^{\rm T}-(m\pm2)\Omega=\kappa \quad {\rm and}\quad
               \omega-m\Omega=-\kappa,   \nonumber \\               
     &&{\rm (c)}:\quad \omega\pm\omega^{\rm T}-(m\pm2)\Omega=-\kappa \quad {\rm and}\quad
               \omega-m\Omega=\kappa,   \nonumber \\
     &&{\rm (d)}:\quad \omega\pm\omega^{\rm T}-(m\pm2)\Omega=-\kappa \quad {\rm and}\quad
               \omega-m\Omega=-\kappa.   
\label{3.8}
\end{eqnarray}

By inspection we see that the radii satisfying condition (a) or condition (d) are not
interesting here, since the excitation condition (\ref{3.7}) is not satisfied in these cases.
If there are radii where condition (b) or (c) is satisfied, the oscillations are excited there since
the excitation condition (\ref{3.7}) is satisfied there.
From condition (b) or (c) we see that such radii are $\pm\omega^{\rm T}\mp2\Omega=2\kappa$.
Since $\omega^{\rm T}$ will be much smaller than $\Omega$ in practical cases, we take as the resonant radius
where oscillations are excited \begin{equation}
     \kappa=\Omega\pm\frac{\omega^{\rm T}}{2},
\label{3.9}
\end{equation}
where $+$ is for the case of $\omega^{\rm T}<0$ and $-$ is for $\omega^{\rm T}>0$, since $\kappa< \Omega$.
In the case where $\omega^{\rm T}$ is much smaller than $\Omega$, this resonance occurs 
at an outer region of the disks, and the frequency ratios of excited oscillations are
roughly 1 : 2 : 3....
Such oscillations will be of interest, but are subjects outside the present issue.

\bigskip\noindent
ii) Vertical resonance of p- or g-mode oscillations

In this case, the radii where resonance occurs efficiently are places where both equations (\ref{3.3})
and (\ref{3.4}) are simultaneously satisfied.
These conditions can be written down in the following four cases:
\begin{eqnarray}
     &&{\rm (a)}:\quad \omega\pm\omega^{\rm T}-(m\pm2)\Omega={\tilde n}^{1/2}\Omega_\bot \quad{\rm and}\quad
               \omega-m\Omega=\kappa,   \nonumber \\
     &&{\rm (b)}:\quad \omega\pm\omega^{\rm T}-(m\pm2)\Omega={\tilde n}^{1/2}\Omega_\bot \quad{\rm and}\quad
               \omega-m\Omega=-\kappa,   \nonumber \\               
     &&{\rm (c)}:\quad \omega\pm\omega^{\rm T}-(m\pm2)\Omega=- {\tilde n}^{1/2}\Omega_\bot \quad {\rm and}\quad
               \omega-m\Omega=\kappa,   \nonumber \\
     &&{\rm (d)}:\quad \omega\pm\omega^{\rm T}-(m\pm2)\Omega=- {\tilde n}^{1/2}\Omega_\bot \quad {\rm and}\quad
               \omega-m\Omega=-\kappa.   
\label{3.10}
\end{eqnarray}

As in the horizontal resonance, we see that cases (a) and (d) are uninteresting even if they have solutions, since 
the excitation condition (\ref{3.7}) is not satisfied.
Resonances resulting from case (b) or case (c), on the other hand, satisfy condition
(\ref{3.7}).
In cases of (b) and (c), the resonant radii are found to be
\begin{equation}
    \kappa=2\Omega-{\tilde n}^{1/2}\Omega_\bot\mp\omega^{\rm T},
\label{3.11}
\end{equation}
where both signs of $\pm$ are possible.
The negative sign is for the case of (b) and positive one is for (c).
Here, $\omega^{\rm T}$ has been assumed to be much smaller than $\Omega$.
This resonant condition is satisfied in the case of ${\tilde n}=2$ and $3$ at inner region of
relativistic disks.

If the resonant radius, $r_{\rm res}$, is determined by solving equation (\ref{3.11}), the frequencies of
resonantly excited oscillations are found to be  
\begin{equation}
    \omega=(m\Omega\pm\kappa)_{\rm res}.
\label{3.11a}
\end{equation}

\bigskip\noindent
iii) Horizontal resonance of vertical p-mode or c-mode oscillations

In this case, the set of equations to be used to determine resonant radii are [see equations
(\ref{3.2}) and (\ref{3.5})]
\begin{eqnarray}
     &&{\rm (a)}:\quad \omega\pm\omega^{\rm T}-(m\pm2)\Omega=\kappa \quad {\rm and}\quad
               \omega-m\Omega=n^{1/2}\Omega_\bot,   \nonumber \\
     &&{\rm (b)}:\quad \omega\pm\omega^{\rm T}-(m\pm2)\Omega=\kappa \quad {\rm and}\quad
               \omega-m\Omega=-n^{1/2}\Omega_\bot,   \nonumber \\               
     &&{\rm (c)}:\quad \omega\pm\omega^{\rm T}-(m\pm2)\Omega=-\kappa \quad {\rm and}\quad
               \omega-m\Omega=n^{1/2}\Omega_\bot,   \nonumber \\
     &&{\rm (d)}:\quad \omega\pm\omega^{\rm T}-(m\pm2)\Omega=-\kappa \quad {\rm and}\quad
               \omega-m\Omega=-n^{1/2}\Omega_\bot.   
\label{3.12}
\end{eqnarray}
As in the previous studies of i) and ii), cases (a) and (d) are outside of our present interest, 
since the oscillations that satisfy the conditions do not satisfy the excitation condition (\ref{3.7}).
In the cases of (b) and (c), on the other hand, oscillations are excited, and the radius is 
characterized by
\begin{equation}
    \kappa=2\Omega-n^{1/2}\Omega_\bot\pm\omega^{\rm T}.
\label{3.13}
\end{equation}
This expression for resonant radii is the same as relation (\ref{3.11}), except that ${\tilde n}$ in relation
(\ref{3.11}) is now replaced by $n$.

The frequencies of resonantly excited oscillations are
\begin{equation}
      \omega=(m\Omega\pm n^{1/2}\Omega_\bot)_{\rm res},
\label{3.14}
\end{equation}
where the resonant radius, $r_{\rm res}$, is now determined by equation (\ref{3.13}).

Finally, it is noted that there is no resonance characterized by the set of equaions (\ref{3.3}) 
and (\ref{3.5}), as far as $n$ and ${\tilde n}$ are moderate integers.

\section{Frequency Variation by Change of Vertical Disk Structure}

Among resonantly excited oscillations discussed in the previous section, the p- and g-mode oscillations
resulting from vertical resonance [i.e., the set described by equations (\ref{3.11}) and (\ref{3.11a})]
are of interest in relation to the observed kHz QPOs,
since their frequencies are in a reasonable frequency range, as discussed in previous papers.

If we want to account for the observed frequency variation of kHz QPOs by the model described by
equations (\ref{3.11}) and (\ref{3.11a}), the frequency of two-armed pattern, $\omega^{\rm T}$,
must change with time.
Variation of $\omega^{\rm T}$ changes the resonant radius [see equation (\ref{3.11})] and thus the
frequencies of oscillations [see equation (\ref{3.11a})] are changed.
This may be one of possible causes of frequency changes of the observed kHz QPOs, but it is not
clear whether a large variation of $\omega^{\rm T}$ required to explain the time variation of
kHz QPOs is generally expected.

Here, one of another possibilities of time variation of resonant radius is considered.
So far, we have assumed that the disk is vertically isothermal and oscillations also
occur isothermally.
Now we relax this assumption, and consider the case where the pressure, $p$, and density, $\rho$, 
are distributed in the vertical direction with a polytropi relation, i.e.,
$p=K\rho^{1+1/N}$, and the polytropic index $N$ changes with time.
It is noted that in the polytropic disks the vertical integration of the vertical hydrostatic balance gives
\begin{equation}
   T_0(r,z)=T_{00}(r)\biggr(1-\frac{z^2}{H^2}\biggr),
\label{4.1a}
\end{equation}
\begin{equation}
   \rho_0(r,z)=\rho_{00}(r)\biggr(1-\frac{z^2}{H^2}\biggr)^N,
\label{4.1b}
\end{equation}
\begin{equation}
   p_0(r,z)=T_{00}(r)\biggr(1-\frac{z^2}{H^2}\biggr)^{1+N},
\label{4.1c}
\end{equation}
where subscript 0 represents the quantities in the equilibrium state and 00 are those on the equatorial plane
(e.g., Kato et al. 2008).

We consider adiabatic disk oscillations on such polytropic disks, assuming that the ratio of the specific 
heat, $\gamma$, is related to $N$ by $\gamma=1+1/N$.
Compared with the case of isothermal disks, derivation of dispersion relation of oscillations in such
polytropic disks are much complicated (e.g., Perez et al. 1997; Silbergleit et al. 2001).
However, in the limiting case  where the couplings between horizontal and vertical motions in an
oscillation mode are neglected, we can easily see that the eigen-frequency of local vertical oscillations with
($\omega$, $m$, $n$) is given by
\begin{equation}
      (\omega-m\Omega)^2-\Psi_n\Omega_\bot^2=0,
\label{4.2}
\end{equation}
where for the fundamental ($n=1$), the first overtone ($n=2$) and the second overtone ($n=3$), we have
(e.g., Kato 2005)
\begin{eqnarray}
      &&\Psi_1=1, \nonumber \\
      &&\Psi_2=2+\frac{1}{N}=1+\gamma,  \nonumber \\
      &&\Psi_3=3+\frac{3}{N}=3\gamma.
\label{4.3}
\end{eqnarray}
It is noted that in the case of isothermal disks, $1/N=0$ and $\Psi_n=n$.

The above consideration suggests that the condition of vertical resonance, equation (\ref{3.3}), is now changed to
\begin{equation}
    [\omega\pm\omega^{\rm T}-(m\pm 2)\Omega]^2-\Psi_n\Omega_\bot^2=0.
\label{4.4}
\end{equation}
This relation shows that the resonant radius of the vertical resonance depends not only on $\omega^{\rm T}$
but also on $\gamma$.
In real accretion disks, a change of mass accretion rate, for example, may bring about a change of
the disk vertical structure.
This change gives rise to a change of radius of the vertical resonance, leading to frequency change
of resonantly excited disk oscillations.

The next subject to be investigated is how the excitation condition (\ref{3.7}) 
is changed in the present case of polytropic disks.
Detailed investigation of this is very complicated in mathematical treatment, since normal mode analyses of
oscillations are troublesome in the case of polytropic disks.
Let us first consider the separability of variables associated with oscillations.
If the radial variation of oscillations is strong, a physical quantity associated with the oscillations, 
say $V(r,z)$, is approximately separated as $V_r(r)V_z(z)$.
In the case of isothermal disks, $V_z(z)$ is the Hermite polynomials [see equation (\ref{2.4})], while
it is the Gezenbauer polynomials in the case of polytropic disks [see Perez et al. 1997; Silbergleit et al. 2001].
This makes description of oscillations complicated.

Furthermore, when we want to describe the non-linear couplings between oscillations, 
we must expand a product of Gezenbauer polynomials into a series of the Gezenbauer polynomials, and must use 
orthogonal relations among the polynomials to separate oscillation modes.
These are much complicated compared with in the case of the Hermite polynomials.
These make a detailed examination of excitation condition troublesome, compared with in the case of
isothermal disks.
Hence, a detailed derivation of excitation condition in the case of polytropic disks is beyond 
the scope of this paper.
Here, we must be satisfied with a rough and physical considerations.
That is, we assume that the excitation condition (\ref{3.7}) still hold even in the case of polytropic disks, 
since it has a simple and reasonable physical meaning that will be free from a particular disk structure,
as mentioned before.

Hereafter we focus our attention only on the case where p- and g-mode oscillations are excited by
the vertical resonance.
In this case the basic equations to be used to specify the resonant radius are the set of equations 
(\ref{4.4}) and (\ref{3.11a}), and the equation to be used to judge whether the oscillations are really excited 
is equation (\ref{3.7}).

By generalizing the procedures in ii) in subsection 3.3,
we easily see that the resonant radii, $r_{\rm res}$, in the cases where oscillations are really excited
are obtained by solving 
\begin{equation}
    \kappa=2\Omega-\Psi_{{\tilde n}}^{1/2}\Omega_\bot \mp\omega^{\rm T},
\label{4.5}
\end{equation}
and the frequencies of the oscillations, $\omega$, are given by
\begin{equation}
     \omega=(m\Omega\pm\kappa)_{\rm res}.
\label{4.6}
\end{equation}

\section {Numerical Results}   

As is shown in equation (\ref{4.6}), the frequencies of oscillations excited at the resonant radius are a 
discrete set, characterized by $m$ and $\pm$.
As the typical frequencies, we take here $(\Omega-\kappa)_{\rm res}$ and $(2\Omega-\kappa)_{\rm res}$
as in the previous papers.
The former oscillation has $m=1$, while the latter does $m=2$.
We take a picture that the observed QPOs come from high energy photons that are Comptonized in a
corona surrounding a geometrically thin disk where the oscillations are generated.
If this picture is adopted, the oscillations with $m=1$ are observed in the twohold frequency
(Kato and Fukue 2006).
Based on this situation, we consider that $2(\Omega-\kappa)_{\rm res}$ and $(2\Omega-\kappa)_{\rm res}$
correspond to the typical twin frequencies of the observed kHz QPOs, and denote them as
\begin{equation}
     2\omega_{\rm LL}=2(\Omega-\kappa)_{\rm res}, \quad
     \omega_{\rm L}=(2\Omega-\kappa)_{\rm res}.
\label{5.1}
\end{equation}

The figures of frequencies $2\omega_{\rm LL}$ and $\omega_{\rm L}$ depend on the resonant radius, 
$r_{\rm res}$, but the relation between $2\omega_{\rm LL}$ and $\omega_{\rm L}$ is free from
detailed models determining $r_{\rm res}$.
That is, the relation depends only on the mass of the central object, $M$, and the spin parameter, $a_*$, 
representing the metric.
The variations of such parameters as $\gamma$ and $\omega^{\rm T}$ determine only 
the allowed range of variation on the $\omega_{\rm L}$ - $2\omega_{\rm LL}$ curve.
The $\omega_{\rm L}$ - $2\omega_{\rm LL}$ relation is shown in figure 1 for some sets of $M$ and $a_*$.
On this figure, the diagram showing the observed relation between the upper and lower frequencies 
of the twin kHz QPOs of some typical sources has been superposed,
assuming that the lower and upper kHz QPOs correspond, respectively, to $2\omega_{\rm LL}$ and $\omega_{\rm L}$.
In the region of $2\omega_{\rm LL} > 600$ Hz, the $\omega_{\rm L}$ - $2\omega_{\rm LL}$ relation
seems to well describe observations if $M=2.4 M_\odot$ and $a_*=0$ are adopted.

%---------------------- Figure 1 -----------------------------------
\begin{figure}
\begin{center}
    \FigureFile(80mm,80mm){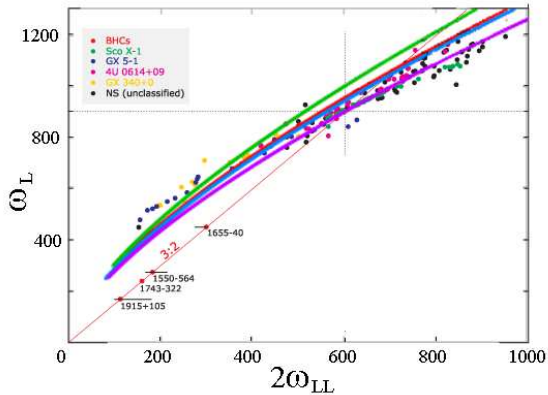}
    %%% \FigureFile(width,height){filename}
\end{center}
\caption{Diagram showing the relation between $\omega_{\rm L}$ and $2\omega_{\rm LL}$ for some values of
mass, $M$, and spin parameter, $a_*$.
The set of $M$ (in units of $M_\odot$) and $a_*$ adopted to draw the curves are, 
from the uppermost to the lowermost curves, (2.0, 0.2)(green), (2.0, 0.0)(red),
(2.4, 0.2)(blue), and (2.4, 0.0)(violet).
Diagram showing the frequency correlation of the twin kHz QPOs of some neutron-star X-ray 
sources (taken from Abramowicz 2005) are superposed by assuming that the upper and lower
kHz QPOs correspond, respectively, to $\omega_{\rm L}$ and $2\omega_{\rm LL}$.   
} 
\end{figure}
%-------------------------------------------------------------------------------------------
Next, we examine which part of the $\omega_{\rm L}$ - $2\omega_{\rm LL}$ curve is allowed 
in the case where $\gamma$ and $\omega^{\rm T}$ vary in resonable ranges.
As a preparation to this study, we first examine how the resonant radius varies by changes of 
$\gamma$ and $\omega^{\rm T}$.
We consider first the case of $\omega^{\rm T}=0$ and ${\tilde n}=2$.
From equation (\ref{4.5}), we see that in this case the resonant radius is described by
\begin{equation}
     \kappa=2\Omega-(1+\gamma)^{1/2}\Omega_\bot.
\label{5.2}
\end{equation}
Figure 2 shows how the resonant radius described by equation (\ref{5.2}) changes as a function of $\gamma$
for two cases of $a_*=0$ and $a_*=0.3$.
The change of resonant radius by change of $\omega^{\rm T}$ is shown in figure 3 by adopting the positive 
sign in equation (\ref{4.5}):
\begin{equation}
    \kappa=2\Omega-(1+\gamma)^{1/2}\Omega_\bot+\omega^{\rm T}.
\label{5.3}
\end{equation}
In figure 3, $\gamma=4/3$ has been adopted and  
two cases of $a_*=0$ and $a_*=0.3$ are shown.
The $r_{\rm res}$ - $\omega^{\rm T}$ relation in the case of $\kappa=2\Omega-(1+\gamma)^{1/2}-\omega^{\rm T}$ is 
obtained by just changing the sign of $\omega^{\rm T}$ in figure 3.
%-------------------- Figure 2 ----------------------------------
\begin{figure}
  \begin{center}
    \FigureFile(70mm,70mm){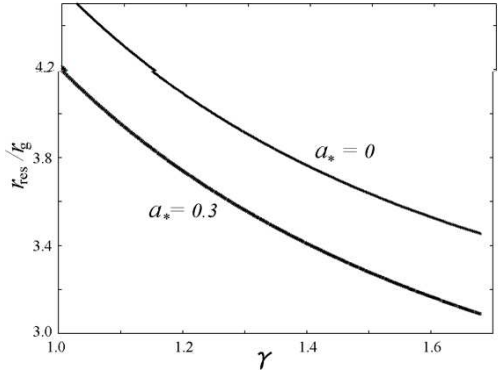}
    %%% \FigureFile(width,height){filename}
  \end{center}
 \caption{$r_{\rm res}/r_{\rm g}$ - $\gamma$ relation obtained by solving equation (\ref{5.2}).
 $\omega^{\rm T}=0$ has been adopted.
 Two cases of $a_*=0$ and $a_*=0.3$ are shown.
} 
\end{figure}
% ------------------ Figure 3 ---------------------------------
\begin{figure}
  \begin{center}
    \FigureFile(70mm,70mm){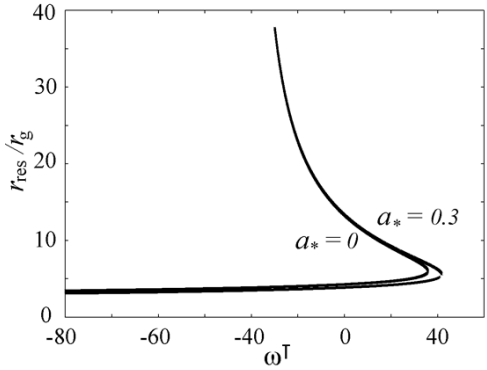}
    %%% \FigureFile(width,height){filename}
  \end{center}
 \caption{$r_{\rm res}/r_{\rm g}$ - $\omega^{\rm T}$ relation obtained by solving equation (\ref{5.3}).
 Two cases of $a_*=0$ and $a_*=0.3$ are shown with $\gamma=4/3$.
} 
\end{figure}
%-------------------------------------------------------

We next examine how much $2\omega_{\rm LL}$ and $\omega_{\rm L}$ vary, when $\omega^{\rm T}$ and $\gamma$ 
are changed in reasonable ranges.
Two cases of $\gamma=1$ and $\gamma=4/3$ are considered, and for each case of $\gamma$, 
$\omega^{\rm T}$ is changed from 0 to -80 Hz (retrograde precession of deformation), 
with $M=2.4\ M_\odot$ and $a_*=0.0$.
The results are shown in figure 4, where $\omega_{\rm L}$ and $\omega^{\rm T}$
are shown as functions of $2\omega_{\rm LL}$. 
In the Z-sources, the frequency of the horizontal branch oscillation (HBO) is known to change with correlation with 
the frequencies of kHz QPOs, i.e., the frequency of HBO is roughly (1/15) of the lower kHz QPO frequency.
So, in figure 4 a straight curve of $(1/15)\times 2\omega_{\rm LL}$ has been added in order to compare
it with the $\vert\omega^{\rm T}\vert$ - $2\omega_{\rm LL}$ relation obtained here.
%---------------------- Figure 4 ----------------------
\begin{figure}
  \begin{center}
    \FigureFile(70mm,70mm){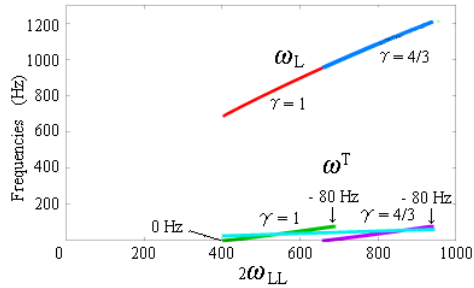}
    %%% \FigureFile(width,height){filename}
  \end{center}
\caption{Diagram showing the $\omega_{\rm L}$ - $2\omega_{\rm LL}$ and $\omega^{\rm T}$ - $2\omega_{\rm LL}$
relations in the cases where $\gamma=1$ and 4/3, in each case $\omega^{\rm T}$ being changed in the 
range of 0 Hz to - 80 Hz (retrograde precession).
For a comparison, the curve of $(1/15)\times 2\omega_{\rm LL}$ is shown in order to compare with observations, since
the frequency of horizontal branch oscillations and that of the lower kHz QPOs are correlated with
the former being about $1/15$ of the latter.
}
\end{figure}
%------------------------------------------------------

The comparison of the line of $(1/15)\times 2\omega_{\rm LL}$ with the $\vert\omega^{\rm T}\vert$ - $2\omega_{\rm LL}$
curve in figure 4 suggests that if changes of $\omega^{\rm T}$ and $\gamma$ are not independent
but correlated, the present model may describe the observed frequency correlation between kHz
QPOs and the horizontal branch QPOs (see the next section).

\section{Discussion}

Comparison of the present disk oscillation model of QPOs with the observed kHz QPOs suggests that masses of
the central sources are around $2.4M_\odot$, if they have no spin (see figure 1).
If they have spin, a larger mass is required to describe observations by our model.
This mass required by our model seems to be rather large compared with that usually supposed as neutron
star mass, although it is not excluded theoretically.

So far as we assume that $2(\Omega-\kappa)_{\rm res}(\equiv 2\omega_{\rm LL})$ and 
$(2\omega-\kappa)_{\rm res}(\equiv \omega_{\rm L})$ correspond, respectively, to the upper and lower kHz QPOs, 
the above conclusion seems to be robust, since the $2\omega_{\rm LL}$ - $\omega_{\rm L}$ relation 
does not depend on detailed models of resonance.
This is true even in the case of one-armed deformed-disks, as far as $2\omega_{\rm LL}$ and
$\omega_{\rm LL}$ are assumed to correspond to the observed twin kHz QPOs.

There are a few possibilities to evade the above conclusion, since some assumptions and simplifications are
involved in the present model.
First, we have assumed that the angular velocity of disk rotation, $\Omega$, is the Keplerian.
In the central part of the disks, this simplification might be violated, since magnetic field anchored to
the central source may be strong enough to modify the disk rotation.

More importantly, we should remember that our analyses are based on an idea that
QPOs are propagating transient phenomena.
That is, in addition to the condition of vertical resonance, equation (\ref{4.4}), we imposed a
condition that the oscillations that are excited most strongly are those whose radial group velocity
just vanishes at the resonant radius, i.e., equation (\ref{3.4}).
Combining these two relations, we have obtained the resonant radius, equation (\ref{4.5}).
The imosed condition (\ref{3.4}), however, is not always clear whether it is most relevant to specify
the resonantly excited oscillations.
If this condition is relaxed, the resonant radius given by equation (\ref{4.5}) is modified.
For example, the resonant radius may differ for each oscillation mode, and $2\omega_{\rm LL}$ - $\omega_{\rm L}$
relation is changed.
Furthermore, we should notice that if the kHz QPOs are trapped oscillations, their frequencies are 
determined by a trapping condition, rather than by vanishing of group velocity.
Then, a different approach from the present paper is necessary to consider frequencies of
QPOs (see Ferreira and Ogilvie 2008; Oktariani et al. 2009).

In this paper we considered the case of ${\tilde n}=2$, i.e., the vertical component of displacement vector
associated with the intermediate oscillations has two nodes in the vertical direction.
In the case of ${\tilde n}=0$ there is no vertical resonance, and in the case of ${\tilde n}=1$,
the frequency of the vertical oscillation and that of the vertical resonance 
are always $\Omega_\bot$, independent of the vertical structure of disks.
Hence, ${\tilde n}=2$ is the possible smallest value of ${\tilde n}$ where the vertical disk structure
can affect the frequency of resonant oscillations.
The intermediate oscillations with ${\tilde n}=2$ are realized when 
the set of ($n$, $n^{\rm T}$) is (0,2), (1,1), (1,3), or (2,0),... in the case of vertically 
isothermal disks.
We assume that the situation does not change much even in the case of polytropic disks.
Then, if we remember that we are now treating the deformation with $m^{\rm T}=2$, the sets with $n^{\rm T}=2$ or
$n^{\rm T}=3$ are of interest among the above sets of ($n$, $n^{\rm T}$), since in the cases of ($m^{\rm T}=2$, 
$n^{\rm T}=2$) and  ($m^{\rm T}=2$, $n^{\rm T}=3$), we can expect low 
frequency deformation of disks by the following reason.

Silbergleit et al. (2001) examined global adiabatic disk oscillations in
polytropic disks, corresponding general considerations of the process deriving relations (\ref{4.3}).
By separating approximately a disturbance associated with disk oscillation, say $V(r,z)$, into a separated form, 
say $V_r(r)V_z(z)$, they solved the resulting wave equation by WKB methods.
They show that the frequency of vertical oscillation,
say $\omega$, is given by
\begin{equation}
     (\omega-m\Omega)^2=\frac{1}{2}n[\gamma(n-1)+(3-n)]\Omega_\bot^2,
\label{6.1}
\end{equation}
where $\gamma$ is the ratio of specific heats, $m$ is the azimuthal wavenumber and $n=0,1,2,...$ 
is an integer characterizing the node number in the vertical direction.
In the cases of ($m=2$, $n=2$) and ($m=2$, $n=3$),
the above equation has solutions of the forms [see equation (\ref{4.3})]:
\begin{eqnarray}
   && \omega= -(\gamma+1)^{1/2}\Omega_\bot+2\Omega \quad (n=2) \nonumber \\
   && \omega=-(3\gamma)^{1/2}\Omega_\bot +2\Omega   \quad \quad (n=3).
\label{6.2}
\end{eqnarray}

Here, the above results are applied to disk deformation, and thus
$\omega$ and $n$ in equation (\ref{6.2}) are regarded, respectively, as
$\omega^{\rm T}$ and $n^{\rm T}$.
Equation (\ref{6.2}) then suggests that a low frequency disk deformation is expected when $n^{\rm T}=2$ and
$n^{\rm T}=3$ for relevant figures of $\gamma$.
For example, in the case of $n^{\rm T}=3$, a steady deformation of disks is possible for $\gamma=4/3$,
and $\omega^{\rm T}$ is negative (retrograde precession) for $\gamma>4/3$,
$\vert\omega^{\rm T}\vert$ increasing with increase of $\gamma$.
This $\gamma$ - dependence of $\vert\omega^{\rm T}\vert$ is qualitatively the same as 
the $\omega^{\rm T}$ - $\gamma$ relation required to describe observational correlation between
kHz QPOs and HBOs.
For example, let us consider the case where $\omega^{\rm T}$ is related to $\gamma$ as
$\omega^{\rm T}=-100(\gamma-0.5)$ (retrograde precession), and $\gamma$ changes 
in the range of $\gamma=2/3$ to $\gamma=4/3$.\footnote{
We do not insist that $\omega^{\rm T}$ and $\gamma$ should be correlated in this way.
This is just an example.
}
The $2\omega_{\rm LL}$ - $\omega_{\rm L}$ and $\vert\omega^{\rm T}\vert$ - $\omega_{\rm L}$ relations in this case 
are shown in figure 5.
On this figure, observed QPOs data for typical sources are superposed, assuming that $\omega_{\rm L}$
corresponds to the upper kHz QPOs.

Finally, we should emphasize that in our QPO model, non-linear couplings and resonances that are considered are
between disk deformation and oscillations.
Concerning oscillations themselves, however, our resonant model is linear;
non-linear and resonant processes among oscillations themselves are not considered.
If we want to describe the fact that the observed amplitudes of neutron-star twin QPOs changes sign as 
the observed frequency ratio of the QPOs passes through the value 3 : 2 (T\"{o}r\"{o}k 2009),
non-linear resonant processes between twin QPOs should be considered as Horak et al. (2009) 
did and succeeded to describe it.
The non-linear resonant processes among oscillations, however, are not considered in our resonant model, since
in our model they are not main processes for determining the oscillations excited and their frequencies.
 
%---------------------------- Figure 5 -------------------------
\begin{figure}
  \begin{center}
    \FigureFile(80mm,80mm){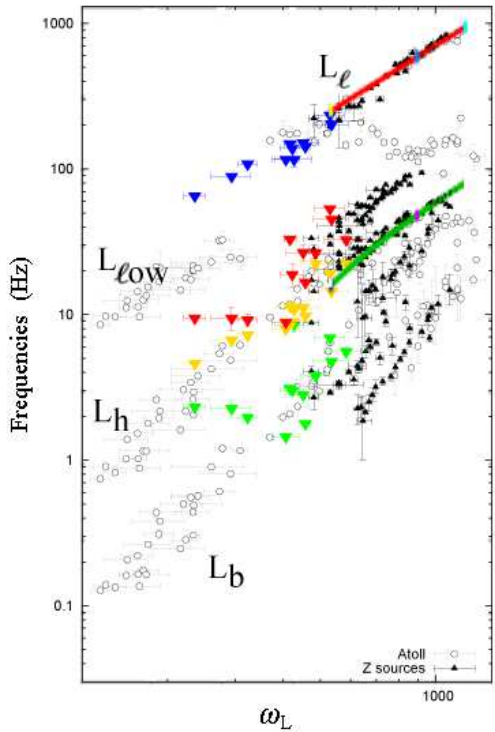}
    %%% \FigureFile(width,height){filename}
  \end{center}
 \caption{Diagram showing $2\omega_{\rm LL}$ - $\omega_{\rm L}$ and $\vert\omega^{\rm T}\vert$ -
 $\omega_{\rm L}$ relations in the case where $\omega^{\rm T}$ and $\gamma$ is related as 
 $\omega^{\rm T}= - 100 (\gamma-0.5)$ and $\gamma$ varies in the range of $2/3$ to $4/3$.
 The mass of the central source and the spin parameters are, respectively, $M=2.4M_\odot$ and $a_*=0$.
 The diagram has been superposed by the diagram showing the observed frequency correlations among 
 kHz QPOs and low frequency QPOs (taken from Boutloukos et al. 2006), assuming that $\omega_{\rm L}$
 corresponds to the higher kHz QPO.
} 
\end{figure}
%-----------------------------------------------------------------
\bigskip
\leftskip=20pt
\parindent=-20pt
\par
{\bf References}
\par
Abramowicz, M. A., \& Klu{\' z}niak, W. 2001, A\&A, 374, L19 \par
Abramowicz, M.A. 2005, Astron. Nachr. 326, No.9 \par
Abramowicz, M.A., Bulik, T., Bursa, M., \& Klu\'{z}niak, W.  2003, A\&A, 404, L21 \par
%Belloni, T., Psaltis, D., van der Klis, M. 2002, ApJ., 572, 392 \par
Boutloukos, S., van der Klis, M., Altamirano, D., Klein-Wolt, M., Wijnands, R., 
      Jonker, P.G., Fender, R.P. 2006, ApJ, 653, 1435 \par
%Bursa, M. 2003, unpublished \par 
Ferreira, B.T. \& Ogilvie, G.I. 2008, MNRAS, 386, 2297 \par
%Hirose, M., Osaki, Y. 1990, PASJ 42, 135\par
%Homan, J. Klein-Wolt, M., Rossi, S. Miller, J.M., Wijnands, R., Belloni, 
%       T., van der Klis, M., Lewin, W.H.G., 2003, ApJ, 586, 1262 \par
Horak, J., Abramowicz, M.A., Klu\'{z}niak, W., Rebusco, P., \& T\"{o}r\"{o}k, G. 2009, A\&A, 499, 536 \par
Kato, S. 2001, PASJ, 53, 1\par 
%Kato, S. 2003a, PASJ, 55, 257 \par
%Kato, S. 2003, PASJ, 55, 801\par
Kato, S. 2004, PASJ, 56, 559 \par
%Kato, S. 2004b, PASJ, 56, 905\par
%Kato, S. 2004c, PASJ, 56, L25\par
%Kato, S. 2005a, PASJ, 57, L17 \par
%Kato, S. 2005b, PASJ, 57, 679 \par
Kato, S. 2005, PASJ, 57, 699 \par
Kato, S. 2008a, PASJ, 60, 111 \par
Kato, S. 2008b, PASJ, 60, 889 \par
Kato, S. 2008c, PASJ, 60, 1387 \par
Kato, S., Fukue, J. 2006, PASJ, 58, 909\par
Kato, S., Fukue, J., \& Mineshige, S. 2008, Black-Hole Accretion Disks --- Towards a New paradigm --- 
  (Kyoto: Kyoto University Press)\par
%Kato, S., Tosa, M. 1994, PASJ, 46, 559 \par

%Klu{\' z}niak, W., \& Abramowicz, M. 2001, Acta Phys. Pol. B32, 3605   \par
%Klu{\' z}niak, W. 2005, Astron. Nachr., 326, 820 \par 
%Klu{\' z}niak, W., Abramowicz, M. A., Kato, S., Lee, W. H., \& Stergioulas,
%   N. 2004, ApJ, 603, L89 \par 
%Klu{\' z}niak, W., Lasota, J-P., \& Abramowicz, M.A. 2005, 
%      astro-ph 0503151\par
%Lamb, F. K. \& Miller, M. C. 2003, astro-ph/0308179 \par
%Li, L.-X., Goodman, J., Narayan, R. 2003, ApJ, 593,980 \par
%Lubow, S.H. 1991, ApJ, 381, 259\par
Machida, M. \& Matsumoto, R., 2008, PASJ, 60, 613 \par
%McClintock, J.E., Remillard, R.A. 2005, "Black Hole Binaries", in
%  Compact Stellar X-ray Sources, eds. W.H.G. Lewin and M. van der Klis,
%   Cambridge University Press, Cambridge, in press; astro-ph/0306213 \par
Meheut, H., \& Tagger, M. 2009, Astro-ph. arXiv:0906.4928v1 \par
Okazaki, A.T., Kato, S., \& Fukue, J. 1987, PASJ, 39, 457\par
Oktariani, F., Okazaki, A.T. \& Kato, S. 2009, in preparation \par
Perez, C.A., Silbergleit, A.S., Wagoner, R.V., \& Lehr, D.E. ApJ, 476, 589 \par
%Psaltis, D., Belloni, T., van der Klis, M. 1999, ApJ, 520, 262\par 
%Remillard, R.A. 2005, Astron. Nachr., 326, 804 \par 
%Shafee, R., McClintock, J.E., Narayan, R., Davis, S.W., Li, L.-X.,
%Remillard, R.A. 2005, astro-ph/0508302\par     
Silbergleit, A.S., Wagoner, R., \& Ortega-Rodriguez, M. 2001, ApJ, 548, 335 \par
T\"{o}r\"{o}k, G. 2009, A\&A, 497, 661 \par
%Tosa, M. 1994, ApJ, 426, L81 \par
%Miller, M.C., Lamb, F.K., \& Psaltis, D. 1998, ApJ, 508,791\par
%van der Klis, M. 2000, ARA\&A, 38, 717    \par
%van der Klis, M. 2004, in Compact stellar X-ray sources (Cambridge University Press), 
%   eds. W.H.G. Lewin and M. van der Klis (astro-ph/0410551)    \par
%van der Klis, M., Wijnands, R. A. D., Horne, K., \& Chen, W. 1997, ApJ,
%     481, L97 \par
%Whitehurst, R. 1988, MNRAS 232, 35    \par
\bigskip\bigskip

\end{document}